\documentclass[12pt]{article}

\makeatletter
\newcommand{\lambdabar}{{\mathchoice
  {\smash@bar\textfont\displaystyle{0.25}{1.2}\lambda}
  {\smash@bar\textfont\textstyle{0.25}{1.2}\lambda}
  {\smash@bar\scriptfont\scriptstyle{0.25}{1.2}\lambda}
  {\smash@bar\scriptscriptfont\scriptscriptstyle{0.25}{1.2}\lambda}
}}
\newcommand{\smash@bar}[4]{%
  \smash{\rlap{\raisebox{-#3\fontdimen5#10}{$\m@th#2\mkern#4mu\mathchar'26$}}}%
}
\makeatother

\RequirePackage[a4paper,top=30.6mm,bottom=38.6mm,left=26mm,right=26mm,footskip=1.3cm]{geometry}
\usepackage{setspace}
\usepackage{amsmath}
\usepackage{amsfonts} 
\usepackage{amssymb}
\usepackage{graphicx}
\usepackage{hyperref}
\usepackage{tensor}
\usepackage{siunitx}
\usepackage{float}
\usepackage{bm}
\usepackage{slashed}
\usepackage{booktabs}
\usepackage{cancel}
\usepackage{multirow}
\usepackage{comment}
\usepackage{bbold}
\usepackage{caption}
\usepackage{amsmath}
\usepackage{enumerate}
\usepackage{cite}
\usepackage{tensor}
\usepackage{slashed}
\usepackage{inputenc}
\usepackage{rotating}
\usepackage{bigfoot}
\usepackage{cancel}
\usepackage{mathtools}
\usepackage{xcolor}
\usepackage{color, colortbl}

\usepackage[textsize=footnotesize]{todonotes}

\numberwithin{equation}{section}

\def \< {\left<}
\def \> {\right>}

\newcommand{\be}{\begin{equation}} \newcommand{\ee}{\end{equation}}
\newcommand{\bea}{\begin{eqnarray}}  \newcommand{\eea}{\end{eqnarray}}
\newcommand{\nn}{\nonumber}

\usepackage{amsfonts, amsthm}
\usepackage[english]{babel}
\usepackage{slashed}
\usepackage{mathrsfs}
\usepackage{amssymb}
\usepackage{color}

\def\Im{{\mathrm{Im}}}

\begin{document}
		
	\begin{center}        
		\Large Instanton Interactions and the Axion Weak Gravity Conjecture
		\end{center}
	
	\vspace{0.4cm}
	\begin{center}        
		{\large Jarod Hattab \;and\;  Eran Palti}
	\end{center}
	
	\vspace{0.15cm}
	\begin{center}  
		\emph{Department of Physics, Ben-Gurion University of the Negev,}\\
		\emph{Be'er-Sheva 84105, Israel}\\[.3cm]

	\end{center}
	
	\vspace{1cm}
	
	
	\begin{abstract}
	\noindent  
	The Weak Gravity Conjecture admits a formulation in terms of long-range charged particle self-interactions being repulsive. We adopt a similar approach to the Axionic Weak Gravity Conjecture by considering long-range instanton interactions. At leading order, in the instanton separation, such interactions are mediated purely by massless scalar fields. Gravitational and gauge interactions are sub-leading. We formulate a repulsive interaction Axionic Weak Gravity Conjecture, which forms a bound on the scalar field dependence of the complex Euclidean instanton action. We show that in standard ultraviolet complete examples, this bound implies the usual formulation of the Axionic Weak Gravity Conjecture, while generally being stronger. Because this formulation is purely a statement about massless scalar fields, it implies that in the absence of massless scalars the usual bound on the axion decay constant could be modified.
    \end{abstract}
	
	\thispagestyle{empty}
	\clearpage
	
	\tableofcontents

\section{Introduction}
\label{sec:int}

The Weak Gravity Conjecture (WGC) \cite{Arkani-Hamed:2006emk} postulates that, in an effective field theory containing a $U(1)$ gauge symmetry, with gauge coupling $g$ , admitting a UV completion that includes quantum gravity, there must exist at least one charged state whose charge-to-mass ratio, $q/m$,  is sufficiently large in $d$-dimensional Planck units $M_P^d$ ,
\begin{equation}
    \left(\frac{d-3}{d-2}\right)m^2\leq q^2g^2(M_P^d)^{d-2}\;.
\end{equation}
Its original bottom-up motivation stands from requiring that black holes should be able to discharge, thereby avoiding stable remnants. The conjecture has been tested in numerous string theory constructions, see \cite{Palti:2019pca,vanBeest:2021lhn,Palti:2020mwc,Harlow:2022ich} for reviews. 

Similar statements are expected to hold for general p-form gauge symmetries \cite{Arkani-Hamed:2006emk,Heidenreich:2015nta} constraining the charge-to-tension ratio $q_p/T_p$ of higher dimensional objects
\begin{equation}
\label{pformnod}
    \frac{p(d-p-2)}{d-2}T_p^2 \leq q_p^2g_p^2 \left(M_P^d\right)^{d-2}\;.
\end{equation}
In the case $p=0$, corresponding to an axion and an instanton, this constraint trivializes. 
Nonetheless, the axion WGC is commonly stated as the requirement that an axion with decay constant $f$ must couple to an instanton with action $S$ and instanton number $n$ such that \cite{Banks:2003sx,Arkani-Hamed:2006emk}
\begin{equation}
\label{awgcappro}
    1 \lesssim \frac{n^2}{f^2S^2}\left(M_P^d\right)^{d-2}\;,
\end{equation}
where the order-one coefficient is more ambiguous than in the particle case. This conjecture has been studied intensively in string theory, following the initial ideas  \cite{Banks:2003sx,Arkani-Hamed:2006emk}. We refer to \cite{Svrcek:2006yi,Bachlechner:2015qja,Junghans:2015hba,Rudelius:2015xta,Montero:2015ofa,Brown:2015iha,Heidenreich:2015wga} for a selection of early work, and \cite{Palti:2019pca,vanBeest:2021lhn,Harlow:2022ich} for reviews. Recent ideas connecting wormhole physics to the axion weak gravity conjecture \cite{Maldacena:2026jqd,DiUbaldo:2026rly} motivate a sharper bound in terms of a precise coefficient. In this work, we adopt a different perspective on the axion WGC, which also leads to a different sharp bound.

In the presence of massless scalar fields $\Phi^M$ , with field space metric $g_{MN}$ , so with kinetic terms 
\be 
{\cal L}_{\Phi} = -\frac{1}{2}\;g_{MN}\;\partial_\mu\Phi^M\;\partial^\mu\Phi^N\;,
\label{Lphi}
\ee
the WGC admits a proposed refinement \cite{Palti:2017elp}, formulated as a repulsive force condition requiring a charged state whose long-range forces are self-repulsive\footnote{See also \cite{Heidenreich:2019zkl} for follow-up work on this approach.} 
\begin{equation}
\label{rfc}
    \left(\frac{d-3}{d-2}\right)m^2+g^{MN} \left(\partial_M m\right) \left(\partial_N m\right) \leq q^2g^2 \:.
\end{equation}
Here, and henceforth, we work in Planck units $M_P^d=1$, unless explicitly stated.
In this work we adopt a similar interaction-based perspective on the axion WGC, proposing a constraint defined through the long-distance interactions between two identical charged instantons.

In order to calculate the analogue of the repulsive force condition for instantons, it is useful to first exchange the repulsive particle force for a positive binding energy condition. This was indeed proposed as the more general formulation in \cite{Aharony:2021mpc}. The positive binding energy then has a natural generalization to instantons. We consider a Euclidean action, coupled to two instantons with (centre) collective coordinates $x_1$ and $x_2$. We then integrate out all fields to arrive at an effective action $I_{\text{eff}}$ , and calculate the interaction action $I_{\text{int}}$ as the difference in the resulting effective actions of the two-instanton action and the single instanton ones \footnote{This is a general procedure, which can be used also in Lorentzian signature and in other spacetime geometries. For example, it was used to calculate the binding energies in AdS of charged particles in \cite{Fitzpatrick:2011hh,Andriolo:2022hax}. It is also used in field theory instanton interactions, see for example \cite{Callan:1977gz,Schafer:1996wv}.} 
\begin{equation}
    I_{\text{int}} \equiv  I_{\text{eff}}(x_1,x_2)-I_{\text{eff}}(x_1)-I_{\text{eff}}(x_2)\;.
\end{equation}
We then propose that the general precise formulation of the Axionic WGC is that the real part of the interaction is non-negative 
\be 
\mathrm{Re} \;I_{\text{int}} \geq 0 \;.
\label{intAWGCgen}
\ee 
The interaction term depends on the collective coordinates of the instantons, and in flat space we consider taking them far apart (approaching infinite separation distance).\footnote{In ADS space, one should take the separation as approaching the boundary, as in \cite{Fitzpatrick:2011hh,Andriolo:2022hax}.}

The constraint (\ref{intAWGCgen}) can be made more explicit in the presence of massless scalar fields with an action (\ref{Lphi}). This is because the leading interaction between instantons at large distance is a monopole one, and only scalar fields mediate such monopole interactions between instantons. Gravity and gauge fields mediate only dipole or higher interactions. This is in contrast to particles, where scalar, gravity and gauge interactions are on an equal footing at large distances.
Denoting the complex (Euclidean) instanton action $I_{\text{instanton}}$ , with real part $S$ and imaginary part $\Theta$, 
\be 
I_{\text{instanton}} \equiv S + i \;\Theta \;,
\ee  
the constraint (\ref{intAWGCgen}) takes the leading form\footnote{The instanton actions should be evaluated on constant backgrounds for the fields $\partial_{\mu} \Phi^M =0$ .}
\begin{equation}
\label{intAWGC}
    g^{MN} \left(\partial_M S\right) \left(\partial_N S\right) \leq g^{MN}\left(\partial_M\Theta\right)\left( \partial_N \Theta \right)\;.
\end{equation} 
So the axion WGC is a condition purely on the scalar interactions between the instantons! The $\Theta$ part of the action corresponds to a repulsive one, while $S$ is associated to attractive ones. 

To see how our formulation of the axion WGC reduces to (\ref{awgcappro}) in special cases, we can make further restrictions on the scalar fields and instanton actions. We consider splitting the scalar field spectrum into three sectors
\be 
\Phi^M =\left\{\varphi, \phi^i,\theta^A\right\} \;.
\ee 
Here $\varphi$ is a special scalar field direction, $\phi^i$ are generic scalars (or pseudo-scalars), and the $\theta^A$ are special periodic pseudo-scalars under which the instanton is charged, so the associated axions. We consider a factorisation of the field space, so write (\ref{Lphi}) as
\be 
{\cal L}_{\Phi} = -\frac{1}{2}\;\partial_\mu\varphi \;\partial^\mu\varphi -\frac{1}{2}\;g_{ij}\;\partial_\mu\phi^i\;\partial^\mu\phi^j -\frac{1}{2}\;g_{AB}\;\partial_\mu\theta^A\;\partial^\mu\theta^B\;.
\label{Lphispli}
\ee
We take the instanton actions to be of the form
\be 
\label{speform}
S = e^{c\varphi }s(\phi^i) \;, \;\; \Theta =  n_A \theta^A \;,
\ee 
with $s$ being a general function, $n_A$ being integers specifying the instanton charges, and $c$ a constant. Physically, the axions $\theta^A$ mediate repulsive (monopole) interactions, while the other scalars mediate attractive ones. Then (\ref{intAWGC}) takes the form
\be 
c^2 + g^{ij}\left(\partial_j \log S\right)\left(\partial_i \log S\right)\leq \frac{g^{AB}n_A n_B}{S^2}\;.
\label{cconststn2}
\ee 
This form of the bound was proposed already in \cite{Vittmann:2018zkf,Palti:2019pca}. 

Writing the (generalisation) of the axion decay constant as 
\be 
g^{AB}n_A n_B = \frac{n^2}{f^2}\;, 
\ee 
we then obtain
\be 
c^2 \leq \frac{n^2}{f^2S^2} \;,
\label{simformcn}
\ee 
which matches the form (\ref{awgcappro}), while being a sharp bound.
Constraints on the magnitude in string theory of $c$ were studied recently in \cite{Etheredge:2026rio}, showing that indeed they satisfy the proposed sharp bounds of \cite{Maldacena:2026jqd,DiUbaldo:2026rly}. 

It also matches the general formulation in the presence of dilatons for the $p$-form WGC using the superextremality condition for black holes \cite{Heidenreich:2015nta}, generalising (\ref{pformnod}),
\begin{equation}
   \left(\frac{\alpha^2}{2}+ \frac{p(d-p-2)}{d-2}\right)T_p^2 \leq q_p^2g_p^2 \left(M_P^d\right)^{d-2}\;.
   \label{genpfomal}
\end{equation}
Here $\alpha$ is a dilaton coupling parameter, which can be related to $c$ in (\ref{simformcn}) in specific contexts. We can directly set $p=0$ in (\ref{genpfomal}) and obtain a non-trivial constraint. In our formulation, this can be understood as the dilaton attractive instanton interaction being weaker than the axion repulsive one.

Apart from the more general and precise formulation of the axion WGC that we propose (\ref{intAWGC}), perhaps the most striking aspect of the instanton interaction approach is that the constraint trivializes in the absence of massless scalar fields. What replaces the constraint in such a setting is as yet unclear, and we discuss this in section \ref{sec:discussion}. But this raises the fascinating possibility that the axion WGC may be violated in settings with no massless scalar fields (beyond the axion itself). 

The paper is set out as follows. We begin in section \ref{sec:inter} by reproducing the repulsive force condition for particles (\ref{rfc}) from a particle probe calculation and showing how this can naturally be extended to extended objects as well as instantons. In the presence of massless scalar fields, we show that the analogue of the repulsive force condition for instantons reproduces the inequality (\ref{intAWGC}). In section \ref{sec:dim}, we then demonstrate that this definition is consistent with the particle repulsive force condition under dimensional reduction. In section \ref{sec:TypeII}, we show how saturation of this bound is realised in $\mathcal{N} = 2$ four-dimensional supergravity for BPS instantons, and recall the motivation from the string theory c-map. In section \ref{sec:discussion} we summarise and discuss how this picture could be modified when the scalar fields acquire a mass, so that the corresponding interactions become finite ranged and other long range interactions must be reconsidered. 

\section{Interactions and Binding Properties}
\label{sec:inter}

As discussed above, a useful way to reformulate the Weak Gravity Conjecture is in terms of the binding properties of charged objects. In asymptotically flat space, for particles and branes this binding criterion reduces at large separation to a balance of long range forces, with gauge repulsion competing against gravitational and scalar attraction. This point of view is particularly natural for instantons. In this case there is no static potential in the usual sense but the relevant quantity is instead the action of a multi instanton configuration relative to the sum of the actions of its constituents. This long range interaction can be computed systematically by treating the objects as probe sources for the light mediating fields. Take two identical objects which may be particles, branes or instantons, with worldvolume dimension $p$. We can choose to embed the worldvolumes of the two objects in parallel into the spacetime, and denote the remaining (normal) separation coordinates as $x^{i}$. So $i$ ranges over $d-p$ directions. We will drop the $i$ index in denoting the position of the object, so the position of the $a^{\mathrm{th}}$ object is denoted as $x_a$. We couple these objects to a set of fields, which we denote collectively as a vector of fields ${\bf B}$. The Euclidean action can be written as the sum of the contribution of the mediating fields $I_{\text{mediat}}$ and of the two probes $I_{1/2}$
\begin{equation}
    I[{\bf B},x_1,x_2] = I_{\text{mediat}}[{\bf B}]+I_1[{\bf B},x_1]+I_2[{\bf B},x_2]\;.
\end{equation}
The interaction between the two objects is then characterized by the quantity
\begin{equation}
    I_{\text{int}} \equiv  I_{\text{eff}}(x_1,x_2)-I_{\text{eff}}(x_1)-I_{\text{eff}}(x_2)\;,
\end{equation}
where $I_{\text{eff}}(x_a)$ is the effective action obtained from integrating out the mediating fields in the presence of the $a$-th probe. A negative interaction $I_{int}<0$ lowers the Euclidean action of the combined configuration relative to that of the separated constituents therefore signalling binding. In particular, for two static particles extending for a Euclidean time $\beta$, one recovers the potential $V$ between the two
\begin{equation}
   \lim_{\beta\rightarrow+\infty} I_{int}/\beta = V(x_1-x_2)\;.
\end{equation}

\subsection{For particles}
\label{fp}

We first illustrate this for particles. We work in a flat Euclidean background with the two probes separated by a spatial distance $r$ with 
\be 
r = |{\bf x}_1-{\bf x}_2| \;.
\ee  
Here by ${\bf x}_a$ we denote the spatial position of the particles. 
At sufficiently large separation, exchange of massive fields is exponentially suppressed, and the leading long range interaction can be obtained by expanding the mediating action to quadratic order around the background. The relevant part of the $d$-dimensional mediating action $I_d$ is therefore
\begin{align}
\label{theory}
    I_d= \int d^dx\sqrt{g}\Bigg[-\frac{R}{2}+\frac{\tau_{AB}(\Phi)}{4}F^A_{\mu\nu}F^B{}^{\mu\nu} +\frac{1}{2}g_{MN}(\Phi)\partial_\mu\Phi^M\partial^\mu\Phi^N\Bigg]\;,
\end{align}
where $R$ denotes the Ricci scalar, and $F^B_{\mu\nu}$ are field strengths for the $U(1)$ gauge fields $A^B_{\mu}$ with gauge kinetic matrix $\tau_{AB}$ whose inverse we note $\tau^{AB}$. The Euclidean action for the particle probes with worldline $\gamma_a$, coordinates $X^\mu_a$, mass $m(\Phi)$ and charges $q_{B}$ is given by
\begin{equation}
\label{actionp}
    I^p_a = \int_{\gamma_a}m\left(\Phi\left(X_a\right)\right)\sqrt{g_{\mu\nu}dX^\mu_adX^\nu_a}-iq_{B}\int_{\gamma_a}A^B_{\mu} dX^{\mu}_a\;.
\end{equation}
Denote the Euclidean time direction $\tau$ and take the two particles to be fixed $\partial_\tau X_a^\mu = (1,\partial_\tau {\bf x}_a) = (1,{\bf 0})$. Expanding around flat space and a constant scalar background,
\begin{equation}
 g_{\mu\nu} = \delta_{\mu\nu}+h_{\mu\nu}\;,\quad\quad\Phi^M(x) = \bar{\Phi}^M+\chi^M(x)\;,
\end{equation}
where $\delta_{\mu\nu}$ is the flat Euclidean metric, $h_{\mu\nu}$ denotes the metric fluctuation, $\bar\Phi^M$ the constant background value of the scalar fields, and $\chi^M$ their fluctuations, the linearised probe action becomes
\begin{equation}
\label{linearprobe}
    I_{a}^p = \int_{-\beta}^{\beta}d\tau \left[ m(\bar{\Phi})\left(1+\frac{h_{\tau\tau}}{2}\right)+\frac{\partial m}{\partial \Phi^M}\Bigg|_{\bar{\Phi}}\chi^M -i q_{B}A^B_\tau+\cdots\right]\;.
\end{equation}
We have taken the particle worldlines to range from $-\beta$ to $\beta$ in Euclidean time in order to regularize the result. A gauge invariant worldline would require sending $\beta \rightarrow \infty$ where the gauge transformations vanish (or taking the Euclidean time periodic, so a Wilson line). 
The leading interaction term can then be computed by integrating out the quadratic mediating fields fluctuations around our background, or equivalently evaluating the quadratic action on the classical solution sourced by the probes.

Consider a collection of fields ${\bf B}$, and expand them as fluctuations ${\bf \delta B}$ about a background ${\bf \bar{B}}$ , 
\be 
{\bf B} = {\bf \bar{B}} + {\bf \delta B} \;.
\ee 
The probes will act as sources for these fields, which depend on their location, so we can write the current associated to the two probes as 
\be 
{\bf J}(x) = {\bf J}_1(x)+{\bf J}_2(x) \;.
\ee 
Note that here by $x$  we denote the full spacetime coordinates. 
The total leading order Euclidean action is then
\begin{equation}
    \int d^dx \left[\frac{1}{2}{\bf \delta B}^T \cdot K\left({\bf \bar{B}} \right)\cdot {\bf \delta B}+{\bf J}(x)\cdot {\bf \delta B}\right] \;,
\end{equation}
with $K\left({\bf \bar{B}}\right)$ the quadratic kinetic operator. Integrating out the fluctuations ${\bf \delta B}$, as a Gaussian integral, yields the interaction action, after subtraction of the individual self-energies, of
\begin{equation}
\label{iintgenf}
I_{\text{int}} = -\int d^dx\;d^dy \; {\bf J}_1(x)^T \cdot G(x-y) \cdot {\bf J}_2(y)\;+... \;.
\end{equation}
with $G(x-y)$ the propagator associated to the form of $K$. For the relevant fields, scalar, gauge and gravitons, the form of $G(x-y)$ is given in (\ref{Gforms}). 

In our case, for a particle, the vector of fluctuations to be integrated out is composed of the fields in (\ref{linearprobe}), so as 
\be 
\delta{\bf B} = \left\{h_{\mu\nu},A^B_{\mu},\chi^M \right\} \;.
\ee 
These are the gravity, gauge and scalar mediators. 
The currents $J^{\mu\nu}_{(h)}$, $J^{\mu}_{(A),B}$ and $J_{(\chi),M}$ associated to the metric, gauge field and scalar field fluctuation respectively, for a static particle at spatial position $
{\bf x}_a$, can be read off of (\ref{linearprobe}) and take the form\footnote{We use $\mathbb{1}_{[-\beta,\beta]}(\tau)$ to denote a function which is equal to one inside the range $-\beta \leq \tau \leq \beta$, and vanishes outside of it. As discussed above, we should take $\beta \rightarrow \infty$ or $\tau$ periodic, to ensure gauge invariance and the associated current conservation.}
\bea
 J^{\mu\nu}_{(h)}(x) &=& \frac{m(\bar{\Phi})}{2}\; \delta^{(d-1)}({\bf x}-{\bf x}_a)\;\delta^{\mu}_\tau \; \delta^{\nu}_\tau  \;\mathbb{1}_{[-\beta,\beta]}(\tau)\;, \nn\\
 J^{\mu}_{(A),B}(x) &=& -i\;q_B \; \delta^{(d-1)}({\bf x}-{\bf x}_a)\;\delta^{\mu}_\tau \;\mathbb{1}_{[-\beta,\beta]}(\tau)\;, \nn \\
 J_{(\chi),M}(x) &=& \left(\partial_M m\right)|_{\bar{\Phi}} \;\delta^{(d-1)}({\bf x}-{\bf x}_a) \;\mathbb{1}_{[-\beta,\beta]}(\tau)\;.
 \eea
In de Donder gauge for the graviton and Feynman gauge for the vectors, the corresponding propagators are
\bea
	G^{(h)}_{\mu\nu,\rho\sigma}(x-y)&=& 2\left(
\delta_{\mu\rho}\delta_{\nu\sigma}+\delta_{\mu\sigma}\delta_{\nu\rho}-\frac{2}{d-2}\delta_{\mu\nu}\delta_{\rho\sigma}\right)D_d(x-y)\;,\nn \\
G^{(A)\,AB}_{\mu\nu}(x-y)&=&\tau^{AB}\delta_{\mu\nu}D_d(x-y)\;,\nn \\
G^{(\chi)\,MN}(x-y)&=&g^{MN}D_d(x-y)\;,
\label{Gforms}
\eea
with $D_d(x-y)$ given by
\begin{equation}
D_d(x-y) = \int\frac{d^{d}k}{(2\pi)^{d}}\frac{e^{ik\cdot \left(x-y\right)}}{k^2}\underset{d\geq 3}{=} \frac{\Gamma\left(\frac{d-2}{2}\right)}{4\pi^{\frac{d}{2}}\left|x-y\right|^{d-2}}\;.
\end{equation}
Here $\Gamma\left(\frac{d-2}{2}\right)$ denotes the standard $\Gamma$-function.  
The resulting interaction term is therefore given by (\ref{iintgenf}) as
\begin{equation}
\label{intp}
    I_{\text{int}} = \left(\tau^{AB}q_{A}q_{B}- m^2 \left(\frac{d-3}{d-2}\right)-g^{MN} \left(\partial_M m\right)\left(\partial_N m \right)\right)\bigg|_{\bar{\Phi}}\int_{-\beta}^{\beta} d\tau_1\int_{-\beta}^{\beta} d\tau_2\,D_d(\tau_1-\tau_2,r) \;.
\end{equation}
In particular for large Euclidean time $\beta\gg r$  and $d\geq 4$ this matches the classic $1/r^{d-3}$ potential 
\begin{equation}
I_{int} \underset{d\geq 4}{=} 2\beta\frac{\Gamma\left(\frac{d-3}{2}\right) }{4\pi^{\frac{d-1}{2}}r^{d-3}}\left(\tau^{AB}q_{A}q_{B}- m^2 \left(\frac{d-3}{d-2}\right)-g^{MN} \left(\partial_M m\right)\left(\partial_N m \right) \right)\bigg|_{\bar{\Phi}}\;.
\end{equation}
The inequality $I_{\text{int}}\geq 0$ (\ref{intAWGCgen}) is then the multiple $U(1)$ version of the repulsive force condition (\ref{rfc}).

\subsection{For Instantons}

We proceed similarly for instantons still with the same d-dimensional action (\ref{theory}) for the mediating fields. At large distances compared to their characteristic size, the leading interaction between two localized probe instantons is determined by their monopole couplings to massless scalar fields $\Phi^M$, provided such couplings are non-vanishing. Indeed, consider for now coupling the instanton solely to massless scalars $\Phi^M$. For that, we write the most general on-shell complex action for the instanton probe centered at position $x_a\in\mathbb{R}^d$ as
\begin{equation}
I^{\mathrm{inst}}_a[\Phi] = S\left(\Phi(x_a)\right)
+i\;\Theta\left(\Phi(x_a)\right)\;,
\end{equation}
with explicit dependence on the $\Phi^M$ but also possibly on their derivatives. 
Expanding the action around the constant scalar background,
\begin{equation}
\Phi^M(x) = \bar{\Phi}^M+\chi^M(x)\;,
\end{equation}
the relevant leading linearised source term is pointlike in $\mathbb{R}^d$
\begin{equation}
I^{\mathrm{inst}}_a[\Phi]-S\left(\bar{\Phi}\right)
-i\;\Theta\left(\bar{\Phi}\right) = \partial_M(S+i\Theta)\big|_{\bar{\Phi}}\;\chi^M(x_a)+\cdots = \int d^dx \;J_{a,M}(x)\;\chi^M(x)+\cdots
\end{equation}
with 
\be 
J_{a,M}(x) = \delta^{(d)}(x-x_a)\;\partial_M(S+i\Theta)|_{\bar{\Phi}} \;.
\ee
Integrating out the action to quadratic order in $\chi^M$, or equivalently evaluating it on the classical solution sourced by two instantons centred at $x_1$ and $x_2$ ,
yields the leading contribution
\bea
\label{rfci}
I_{\text{int}} &=& -\int d^dx\; d^dy\; J_{1,M}(x)\;g^{MN}D_d(x-y)\;J_{2,N}(y)  +\cdots\;,\\
\label{rfci2}
&\underset{d\geq 3}{=}& -\frac{\Gamma\left(\frac{d-2}{2}\right)}{4\pi^{\frac{d}{2}}\rho^{d-2}}\;g^{MN}\partial_M\left(S+i\Theta\right)\partial_N\left(S+i\Theta\right)\big|_{\bar{\Phi}} +\cdots\;,
\eea
with $\rho$ denoting the instanton separation
\be 
\rho = |x_1-x_2|\;.
\ee  
In particular, requiring positivity of the real part of this interaction term $\mathrm{Re}\; I_{\text{int}}\geq 0$ yields the inequality  (\ref{intAWGC})
\begin{equation}
     g^{MN} \left(\partial_M S\right) \left(\partial_N S\right) \leq g^{MN}\left(\partial_M\Theta\right)\left( \partial_N \Theta \right)\;.
\end{equation}
As discussed in the introduction, we conjecture that this provides the general form of the Weak Gravity Conjecture for instantons coupled to massless scalar fields.

Terms containing spatial derivatives are suppressed because every derivative acting on the propagator generates an additional power of $\frac{1}{\rho}$.
Also nonlinear couplings involving higher powers of $\chi^M$ generate additional powers of $D_d(\rho)$. Consequently, at asymptotically large separation the scalar monopole exchange dominates whenever such a coupling is present.

Similarly, no monopole coupling is possible for gauge fields since a linear term would not be gauge invariant meaning that coupling to gauge fields enter first through a dipole one with 
\begin{equation}
I^{\mathrm{inst}}_a[\Phi^M,A^B_\mu] = S(\Phi(x_a))+i\Theta(\Phi(x_a))+C^{B,\mu\nu}(\Phi(x_a))\;F^B_{\mu\nu}(x_a)+\cdots\;,
\end{equation}
with $C^{B,\mu\nu}(\Phi(x_a))$ some antisymmetric tensorial data carried by the instanton which might depend on the value of the scalar fields. Because $F^B_{\mu\nu}$ involves a spatial derivative, these will contribute to subdominant $1/\rho$ interaction terms. 


The same is true of gravitational interactions. There is no non trivial gravitational monopole coupling; diffeomorphism invariance implies that the first genuine local coupling to gravity involves the Riemann tensor $R_{\mu\nu\rho\sigma}$
\begin{equation}
    I^{\mathrm{inst}}_a[\Phi^M,g_{\mu\nu}]
    =
    S(\Phi(x_a))+i\Theta(\Phi(x_a))
    +D^{\mu\nu\rho\sigma}(\Phi(x_a))\;R_{\mu\nu\rho\sigma}(x_a)+\cdots \;,
\end{equation}
where $D^{\mu\nu\rho\sigma}(\Phi(x_a))$ is  some tensorial data carried by the instanton, leading again to subleading higher multipole interactions. If both instantons couple through curvature operators, the corresponding one graviton exchange contains four derivatives of the propagator thus scaling as $\frac{1}{\rho^{d+2}}$ (up to possible cancellation). Consequently, neither the gauge fields nor the graviton contribute at the same $1/\rho^{d-2}$ order as massless scalar fields. The leading long range interaction between instantons is therefore controlled by the scalar dependence of their action.

\section{Special Cases}
\label{sec:examples}

In this section we study special cases where the repulsive interaction condition for instantons takes the special form (\ref{cconststn2}). In particular, we match this to the repulsive force condition for particles (\ref{rfc}) in two ways. First, by dimensional reduction on a circle, and second by using the c-map in $\mathcal{N}=2$ four-dimensional supergravity arising from string theory. 

One of the central points is to illustrate how the gravitational contribution to the particle self-interaction is indeed always matched to the instanton interaction contribution of a massless gravitationally coupled scalar. In the case of dimensional reduction the scalar is the radius of the circle, while in the string theory case it is the string coupling. 

We note that the connection between the particle and instanton conjectures has been previously studied, through dimensional reduction already in the early works \cite{Banks:2003sx,Arkani-Hamed:2006emk,Rudelius:2015xta,Montero:2015ofa,Brown:2015iha,Heidenreich:2015wga,Heidenreich:2015nta}, and through the c-map in \cite{Hebecker:2015zss,Vittmann:2018zkf,Palti:2019pca}. The key aspect of our analysis is to emphasise the match to the form (\ref{intAWGC}), including the general scalar contributions, and to illustrate the presence of a universal gravitational scalar interaction in these settings.

\subsection{Dimensional Reduction}
\label{sec:dim}

A special case of the identity (\ref{intAWGC}) can be recovered through dimensional reduction. Take the same theory as in section (\ref{fp}) in dimension $D = d+1$ and compactify the theory on $\mathbb{R}^d\times S^1$ with circle coordinate $y\sim y+L$. Introducing the radion \(\varphi(x)\) and the Kaluza-Klein gauge field $B_\mu(x)$ with $\mu\in 1,\cdots, d$, we take the Einstein frame metric ansatz as
\begin{equation}
    ds^2_D = e^{2a\varphi}ds^2_d+e^{2b\varphi}\left(dy+B_{\mu}dx^\mu\right)^2\;,
\end{equation}
with
\begin{equation}
    a = -\frac{1}{\sqrt{2(d-1)(d-2)}}\;,\quad\quad b = \sqrt{\frac{d-2}{2(d-1)}}\;.
\end{equation}
Winding the particles with action (\ref{actionp}) around the $S^1$ at fixed position $x_a\in\mathbb{R}^d$ with winding numbers $w\in\mathbb{Z}$  will yield d-dimensional instantons. Restricting to the Kaluza-Klein zero modes $\Phi_0^M$ of $\Phi^M$ their action reduces to
\begin{equation}
    I^{\mathrm{inst}}_a = |w|\; e^{b\varphi}\;L\; m(\Phi_0^M(x_a))-i\;w\; q_{B}\;\theta^B(x_a)\;,
\end{equation}
in terms of d-dimensional axion fields.
\begin{equation}
    \theta^B = \int_0^{L}A_y^Bdy\;.
\end{equation}
Upon dimensional reduction each term in the action, the Ricci scalar, the gauge fields and scalar fields, give rise to lower dimensional massless scalar fields $\varphi$, $\theta^B$, and $\Phi_0^M$ respectively, appearing in the d-dimensional action as
\begin{equation}
    I_D\rightarrow I_d\supset\frac{1}{2}\int d^dx\sqrt{g}\left(\frac{L \left(M_P^D\right)^{D-2}}{2}(\partial\varphi)^2+\frac{\tau_{AB}\;e^{-2 b\varphi}}{L}\partial_\mu \theta^A\partial^\mu\theta^B+L\;g_{MN}\;\partial_\mu \Phi_0^M\;\partial^\mu \Phi_0^N\right)\;,
\end{equation}
with inverse scalar metrics 
\begin{equation}
    g^{\varphi\varphi} \equiv \frac{2}{L\left(M_P^D\right)^{D-2}},\quad\quad g^{AB} \equiv \tau^{AB} L\;e^{2b \varphi},\quad\quad g_0^{MN} \equiv g^{MN}/L\;.
\end{equation}
If the same interaction calculation is performed in $\mathbb{R}^d\times S^1$ in the same flat background $ds_D^2 = \delta_{\mu\nu}dx^\mu dx^{\nu}+dy^2$, for two particles with spatial separation $r$ and worldlines winding the circle with winding number $w$, it will then yield exactly as in (\ref{intp})
\begin{equation}
\label{Drfc}
    I_{\text{int}} = \left(\tau^{AB}q_{A }q_{B}-\frac{1}{\left(M_P^D\right)^{D-2}} m^2 \frac{D-3}{D-2}-g^{MN}\partial_M m \partial_N m \right)\bigg|_{\bar{\Phi}}w^2 \int_0^{L} dy_1\int_0^{L} dy_2\,D_{d,S^1}(y_1-y_2,r)\;,
\end{equation}
but with instead the massless propagator on $\mathbb{R}^d\times S^1$
\begin{equation}
 D_{d,S^1}(y_1-y_2,r) = \frac{1}{L}\sum_{n\in\mathbb{Z}}\int\frac{d^{d}k}{(2\pi)^{d}}\frac{e^{ik\cdot r}e^{2\pi i n (y_1-y_2)/L}}{k^2+\left(\frac{2\pi n}{L}\right)^2}\;.
\end{equation}
Noting that 
\begin{equation}
 \int_0^{L} dy_1\int_0^{L} dy_2\,D_{d,S^1}(y_1-y_2,r) = LD_d(r)\;,
\end{equation}
and that the background $ds_D^2 = \delta_{\mu\nu}dx^\mu dx^{\nu}+dy^2$ is equivalent to the flat d-dimensional background where we set the vev of the radion to zero $\bar{\varphi} =0$, then expression (\ref{Drfc}) becomes in terms of the dimensionally reduced quantities
\begin{equation}
I_{int} =  -\frac{\Gamma\left(\frac{d-2}{2}\right)}{4\pi^{\frac{d}{2}}r^{d-2}}\;\left(g_0^{MN}\left(\partial_MS\right)\left(\partial_NS\right)+ g^{\varphi\varphi}\left(\partial_\varphi S\right)\left(\partial_\varphi S\right)-g^{AB}\left(\partial_A\Theta\right)\left(\partial_B\Theta\right)\right)\bigg|_{(\bar{\Phi}_0,\bar{\varphi} =0)} \;,
\end{equation}
where
\begin{equation}
    S = |w| e^{b\varphi}L m(\Phi_0),\quad\quad \Theta=-w q_{B}\theta^B\;,
\end{equation}
which reduces to the special case (\ref{speform}) of the general identity (\ref{rfci2}).

A nonzero constant $\varphi$ background can simply be absorbed into a constant rescaling of the lower dimensional Einstein-frame metric, together with a rescaling of the $S^1$ radius.

\subsection{Type II strings and instantons under the c-map}
\label{sec:TypeII}

Similarly, there is a simple way to illustrate this repulsive interaction condition in $\mathcal{N} = 2$ supergravity and relate it to  its particle version under the c-map. Indeed, this was used to propose (\ref{cconststn2}) as a constraint already in \cite{Vittmann:2018zkf,Palti:2019pca}. 

First, let's recall the corresponding structure for charged particles in four-dimensional $\mathcal{N} = 2$ supergravity. See \cite{DallAgata:2011zkh} for a review, and we use the conventions in \cite{Palti:2017elp}. The Lagrangian density for the two-derivative bosonic sector takes the form
\begin{equation}
    \mathcal{L}  = \sqrt{-g}\left[\frac{R}{2}-g_{i\bar{\jmath}}\,\partial_{\mu} s^i\partial^{\mu}\bar{s}^{\bar{\jmath}}+\mathcal{I}_{IJ}F^I_{\mu\nu}F^{J}{}^{\mu\nu}-\frac{1}{2}\mathcal{R}_{IJ}F^I_{\mu\nu}F^{J}_{\rho\sigma}\,\epsilon^{\mu\nu\rho\sigma}-\frac{1}{4}h_{mn}\,\partial_\mu q^{m}\partial^{\mu}q^n\right]\;.
\end{equation}
The indices $I,J = 0,\dots,n_V$ label $n_V+1$ abelian gauge fields with field strength $F^I_{\mu\nu}$ where $n_V$ represents the number of vector multiplets. Here, $\epsilon^{\mu\nu\rho\sigma}$ denotes the Levi-Civita tensor with $\epsilon^{0123} = 1/\sqrt{-g}$. The gauge couplings and theta angles are encoded in the scalar dependent matrices
\begin{equation}
    \mathcal{N}_{IJ}(s) = \mathcal{R}_{IJ}+i\mathcal{I}_{IJ}\;,
\end{equation}
where $\mathcal{I}_{IJ}$ is negative definite for consistency. 
The complex scalars $s^i$ with $i\in 1,\cdots,n_V$ belong to the vector multiplets and parameterize a special Kähler manifold with metric $g_{i\bar{\jmath}}$ , expressed locally in terms of a Kähler potential $K$ ,
\begin{equation}
    g_{i\bar{\jmath}} = \partial_i\partial_{\bar{\jmath}}K,\quad\quad K = -\ln i\left(F_I\bar{X}^I-\bar{F}_IX^I \right)\;,
\end{equation}
whose geometry is encoded in a holomorphic period vector $\Pi(s)=(X^I(s),F_I(s))^T$. In particular, if we note $F_{IJ} = \partial_{I}F_J$ then
\begin{equation}
    \mathcal{N}_{IJ}(s) = \bar{F}_{IJ}+2i\frac{\Im\, F_{IK} \Im\, F_{JL}X^KX^L}{\Im\, F_{MN} X^M X^N}\;.
\end{equation}
The real scalars \(q^m\), with \(m=1,\ldots,4n_H\), arise from the \(n_H\) hypermultiplets and parameterize a quaternionic Kähler manifold with metric $h_{mn}$.

It will also be useful to introduce the positive definite matrix
\begin{equation}
    \mathcal{H} = \begin{pmatrix}
-(\mathcal{I}+\mathcal{R}\mathcal{I}^{-1}\mathcal{R})& \mathcal{R} \mathcal{I}^{-1} \\
\mathcal{I}^{-1}\mathcal{R} & -\mathcal{I}^{-1} 
\end{pmatrix}\;,
\end{equation}
which satisfies for any constant vector $\gamma = (p^{I},q_J)^T$ the identity
\begin{equation}
\label{identity}
 \frac12\gamma^T\mathcal{H}\gamma = |Z_\gamma|^2+g^{i\bar{\jmath}}\left(\nabla_i Z_\gamma\right)\left( \nabla_{\bar{j}}\bar{Z}_{\gamma}
\right)\;,
\end{equation}
where $Z_\gamma$ is the central charge
\begin{equation}
    Z_\gamma = e^{\frac{K}{2}}(\Omega \gamma)^T\cdot \Pi(s)\;,\quad\quad \Omega = \begin{pmatrix}
0 & 1 \\
-1 & 0 
\end{pmatrix}\;,
\end{equation}
and $\nabla_i$ is the Kähler covariant derivative on the vector multiplet moduli space $\nabla_i Z_\gamma = \left(\partial_i+ \frac{1}{2}\partial_i K\right)Z_\gamma$.

In particular, for a BPS state of charge $\gamma = (p^{I},q_J)^T$ the mass is fixed by the central charge
\begin{equation}
M_\gamma = |Z_\gamma|\;,
\end{equation}
and the identity can be interpreted noticing that it can be rewritten
\begin{equation}
\frac{1}{2}  \gamma^T\mathcal{H}\gamma = M_\gamma^2+4g^{i\bar{\jmath}}\left(\partial_i M_\gamma\right) \left(\partial_{\bar{j}}M_\gamma\right)\;,
\end{equation}
as a no force condition saturating the repulsive force condition (\ref{rfc}).

We can ask the same thing of the BPS instantons. We want to find the interaction equality saturated by such BPS objects. We can utilize string theory and consider type II string theories compactified on Calabi-Yau manifolds to four dimensions. The dimensional reduction of the action and instantons is familiar, and here we utilize specifically the results and conventions of \cite{Alexandrov:2021shf}. Working in Type IIA, the instantons correspond to $E2$ branes wrapping three cycles in the Calabi-Yau which couple to the hypermultiplet sector. The sector has $n_H$ hypermultiplets. There is the universal hypermultiplet containing the four dimensional dilaton $\varphi$, the scalar $\sigma$ dual to the four dimensional two form descending from $B_2$ and two axions $\zeta^0,\tilde\zeta_0$, arising from the ten dimensional three form $C_3$. There are then an additional $h^{2,1} = n_H-1$ hypermultiplets, with $h^{2,1}$ being the associated Hodge number, composed of the complex structure moduli $z^a$ and the Ramond-Ramond scalars $\zeta^a,\tilde\zeta_a$. It is convenient to combine the axions into a vector $\zeta = (\zeta^A,\tilde\zeta_A)^T$, with index range $A=\{0,a\}$.

The tree-level moduli space metric $h_{mn}$ then has a line element of the form (see, for example, \cite{Alexandrov:2021shf})
\begin{equation}
 d\varphi^2+4g_{a\bar{b}}\,dz^ad\bar{z}^{\bar{b}}+\frac{e^{-\varphi}}{2}d\zeta^{T}\mathcal{H}_{cs}d\zeta+\frac{e^{-2\varphi}}{16}\left(d\sigma+(\Omega \zeta)^{T}d\zeta\right)^2\;,
\end{equation} 
where $g_{a\bar{b}}$ is the special Kahler metric on the complex structure moduli space and $\mathcal{H}_{cs}(z)$ the corresponding period matrix. We stress that two distinct special Kahler geometries appear here. The quantities $g_{i\bar{\jmath}}(s)$ and $\mathcal{H}(s)$ belong to the vector multiplet moduli space, parametrized in type IIA by the complexified Kahler moduli $s^i$, while $g_{a\bar b}(z)$ and $\mathcal{H}_{cs}(z)$ are associated with the complex structure moduli $z^a$ with associated holomorphic period vector $\Pi_{cs}(z) = (X_{cs}^I(z),F_{cs,\, I}(z))^T$. The latter provide the special Kahler data entering the c-map construction of the hypermultiplet metric. The instantons wrapping three cycles have charges $\gamma$ under the axions $\zeta$ and their actions read (see, for example, \cite{Alexandrov:2021shf})
\begin{equation}
S_\gamma+i\Theta_\gamma = 8\pi e^{\varphi/2}e^{\frac{K_{cs}}{2}}|(\Omega \gamma)^T \cdot\Pi_{cs}(z)|+2\pi i (\Omega \gamma)^T \cdot\zeta\;,
\end{equation}
with $K_{cs}$ the complex-structure Kahler potential. 
The action satisfies
\bea
 2\partial_\varphi\left(S_\gamma+i\Theta_\gamma\right)\partial_\varphi \left(S_\gamma+i\Theta_\gamma\right) &=& \frac{S^2_\gamma}{2}\;,\nn\\
  2g^{a\bar{b}}\partial_{a}\left(S_\gamma+i\Theta_\gamma\right)\partial_{\bar{b}} \left(S_\gamma+i\Theta_\gamma\right) &=& 2g^{a\bar{b}}\left(\partial_a S_\gamma\right) \left(\partial_{\bar{b}}S_\gamma\right)\;,\\
 \left[\partial_{\zeta}\left(S_\gamma+i\Theta_\gamma\right)\right]^T\cdot 4e^{\varphi}\mathcal{H}_{cs}^{-1}\cdot\left[\partial_{\zeta} \left(S_\gamma+i\Theta_\gamma\right)\right] &=& -16\pi^2 e^{\varphi} (\Omega \gamma)^T\mathcal{H}_{cs}^{-1}(\Omega\gamma) = -16\pi^2 e^{\varphi} \gamma^T\mathcal{H}_{cs}\gamma \nn\;,
\eea
and therefore the instanton interaction takes the form
\begin{equation}
 2h^{mn}\partial_m \left(S_\gamma+i\Theta_\gamma\right)\partial_n \left(S_\gamma+i\Theta_\gamma\right) = \frac{S^2_\gamma}{2}+2g^{a\bar{b}}\left(\partial_a S_\gamma\right) \left(\partial_{\bar{b}}S_\gamma\right)-16\pi^2 e^{\varphi} \gamma^T\mathcal{H}_{cs}\gamma  \;.
 \label{sintiiains}
\end{equation}
Because the instanton action is given by the central charge, the interaction (\ref{sintiiains}) vanishes by the identity (\ref{identity}). This can be understood as 
balancing the scalar complex structure $2g^{a\bar{b}}\left(\partial_a S_\gamma\right) \left(\partial_{\bar{b}}S_\gamma\right)$, the dilaton $\frac{S^2_\gamma}{2}$ and the axion $-16\pi^2 e^{\varphi} \gamma^T\mathcal{H}_{cs}\gamma$ interaction terms.

Similar BPS identities studied in \cite{Palti:2017elp}, impose constraints on the spectrum of BPS states. Replacing $\mathcal{N}_{IJ}$ in the definition of $\mathcal{H}$ by $F_{IJ}$ yields \cite{Ceresole:1995ca}
\begin{equation}
\label{identity2}
 \frac12\gamma^T\mathcal{H}(F_{IJ})\gamma = |Z_\gamma|^2-g^{i\bar{\jmath}}\left(\nabla_i Z_\gamma\right) \left(\nabla_{\bar{j}}\bar{Z}_{\gamma}\right)\;.
\end{equation}
Since $\Im(F_{IJ	})$ has $n_V$ strictly positive eigenvalues and one negative, then in the real charge
space $\gamma\in\mathbb{R}^{2n_v+2}$, there are $2n_V$ directions for which the scalar interactions are strictly stronger than the gravitational interaction and two where they act weaker.

This identity applied to the vector multiplet sector in \cite{Palti:2017elp} motivated the conjecture that there necessarily exist a charge directions for which the scalar interactions dominate gravitational ones
\begin{equation}
 g^{MN}\left(\partial_M m\right) \left(\partial_{N}m\right)\geq\frac{1}{2}m^2\;.
\end{equation}
Transposed to instantons this would mean that there always exists an instanton for which the dilaton/radion $\varphi$ interaction is subdominant to the interaction mediated by the remaining scalars $\phi^i$
\begin{equation}
 g^{ij}\left(\partial_i S\right) \left(\partial_{j}S\right)  \geq \left(\partial_\varphi S\right)^2\;.
\end{equation}
Similar identities of the form studied in \cite{Palti:2017elp,DallAgata:2020ino} could impose further constraints on the relative strength of the long-range interactions mediated by the different scalar fields.

\subsection{$M_p^d$ Factors and Gravity Decoupling}
\label{sec:discussion}

It is useful to restore factors of $M_P^d$ to study what becomes of such inequalities in a gravitational decoupling limit. Indeed one expects some form of trivialisation as gravity decouples and $M_P^{d}\rightarrow \infty$. For particles this is explicit in the WGC bound
\begin{equation}
  \left(  \frac{d-3}{d-2} \right) m^2\leq q^2g^2(M_P^d)^{d-2}\;.
\end{equation}
If massless scalar fields are present, this leads to non trivial constraints in the decoupled QFT \cite{Palti:2017elp,Heidenreich:2019zkl}.

Similarly, for instantons, gravity decoupling leads to a trivialisation of the standard inequality
\begin{equation}
S^2\lesssim \frac{n^2}{f^2}\left(M_P^d\right)^{d-2}\;.
\end{equation}
However, from the perspective of the instanton interactions this is somewhat puzzling since there is no graviton exchange contribution to the interaction. The resolution of this is that the constraint is only non-trivial in the presence of at least one additional (beyond the axion) massless scalar, which we can denote the dilaton, and that this scalar must couple with gravitational strength. The gravitational strength coupling, so like a graviton, then ensures the triviality of the constraint in the gravity decoupling limit. 

In terms of the form (\ref{cconststn2}), the gravitational strength coupling is the statement that $c$ has to be of order one in Planck units. So the dilaton coupling to the instanton is controlled by the Planck scale. In the two examples studied in this section, indeed both the circle radius and the string coupling are gravitationally coupled scalars. Note that, as in the particle case, even in the gravity decoupling limit there can still remain a non-trivial constraint involving other non-gravitationally coupled scalars. 

\section{Discussion}
\label{sec:discussion}

In this note we proposed a reformulation of the Axion Weak Gravity Conjecture in terms of long range interactions between instantons (\ref{intAWGCgen}). This was motivated by a similar repulsive interaction proposal for the particle Weak Gravity Conjecture \cite{Arkani-Hamed:2006emk,Palti:2017elp,Heidenreich:2019zkl,Aharony:2021mpc}. Of course, the quantum gravity microscopic physics that could underlie such repulsive requirements remains undetermined as yet, so it is important to emphasise the conjectural nature of the proposal. 

When instantons couple to massless scalar fields $\Phi^M$, with moduli space metric $g_{MN}$, the repulsive interaction condition reduces to the existence of an instanton with complex Euclidean action $I = S(\Phi^M)+i\Theta(\Phi^M)$ such that 
\begin{equation}
 g^{MN}\left(\partial_M S\right)\left(\partial_N S\right)\leq g^{MN}\left(\partial_M \Theta\right)\left(\partial_N \Theta\right)\;.
 \label{gssdef}
\end{equation}
This formulation is a sharp bound, without any order one ambiguities. It is also a very general one.

Perhaps the most striking element of the proposal is that there is no contribution to the constraint coming from gauge or gravitational instanton interactions. The only relevant long-range interactions are those of massless scalar fields. 

In the cases where the (additional to the axion) massless scalar field is a dilaton-type field, as in all known ultraviolet complete examples, the constraint implies (but can be stronger than) the often utilized form (\ref{awgcappro}) ,
\begin{equation}
   f^2 S^2\lesssim \left(M_P^d\right)^{d-2}\;,
\end{equation}
with $f$ the associated axion decay constant.

The interaction approach suggests that a central open question from this perspective is what happens when the scalar fields are massive. It is not clear what replaces the constraint in such cases, and in particular it opens the possibility for its violation. 

At the first level, one can consider what happens when both the repulsive axions and the attractive scalars have a mass. Since for massive scalars (including the axion) the interaction strength falls off exponentially, the leading interactions at large instanton separation would be higher derivative gravitational and gauge interactions (so suppressed by higher powers of the instanton separation). These would appear through couplings to the associated field strengths of the form\footnote{The couplings must be diffeomorphism invariant. So by $C^{B,\,\mu\nu}$ and $D^{\mu\nu\rho\sigma} $ we mean gauge or gravitational quantities evaluated on the instanton background.  For point-like microscopic instantons, such as wrapped D-branes, the gravitational coupling must therefore appear through independently diffeomorphism invariant quantities, such as $R^2$.}
\begin{equation}
I^{\text{inst}} \supset C^{B,\,\mu\nu}F^B_{\mu\nu}+D^{\mu\nu\rho\sigma} R_{\mu\nu\rho\sigma} +\cdots\;.
\end{equation}
One could then imagine a repulsive constraint on such couplings. 

An alternative idea is to consider demanding repulsive interactions at finite separation distances, so that the exponential suppression for the scalars can be avoided (so at distances comparable to the the inverse scalar masses).\footnote{In order for the scalar interactions to dominate over the higher multipole gauge and gravitational ones, the distance scales should be larger than the size of the instanton. This is natural for string theory instantons, but for field theory it may or may not hold.} As long as the mass for all the scalars is below the cutoff of the theory, one could demand the repulsive constraint to hold at the scale appropriate to the most massive field, in which case it should yield something similar to (\ref{gssdef}).  From the perspective of effective field theory, this is the most reasonable expectation, since a small mass for the scalars should not modify the ultraviolet values of the parameters in the theory. 

It is possible to consider a more refined approach, demanding repulsive interactions at various intermediate distance scales. The case when the axions are more massive than the attractive scalars, is in some sense similar to the question of the meaning of the WGC for particles when the gauge fields obtain a small mass. The case when the scalars are more massive than the axions is not possible in the particle setting, since gravity is massless, and so in this sense is more novel. At large distance scales, where the axion interactions are relevant but the scalar interactions are not,  the constraint is trivially satisfied with no bound on the axion decay constant. It would be interesting to understand if this could open new opportunities for model building. 

It is interesting that the wormhole-based approach of \cite{Maldacena:2026jqd,DiUbaldo:2026rly} suggests a constraint should arise even with only a light (or massless) axion. It may be that in such cases, if they arise in an ultraviolet complete theory, there are other obstructions to the wormholes. It would be interesting to understand the compatibility of the two approaches further.

\vspace{0.1cm}
{\bf Acknowledgements}
\noindent
 The work of JH and EP is supported by the Israel Science Foundation (grant No. 1655/26) and by the German Research Foundation through a German-Israeli Project Cooperation (DIP) grant ``Holography and the Swampland". The work of EP is supported by the Israel planning and budgeting committee grant for supporting theoretical high energy physics.

\bibliographystyle{jhep}
\bibliography{WGC}
\end{document}